\documentclass[preprint,review,12pt]{elsarticle}

\usepackage{amssymb}
\usepackage{amsmath}
\usepackage{array}
\usepackage{xcolor}
\usepackage{algorithm}
\usepackage{algpseudocode}

\usepackage{etoolbox}
\AtBeginEnvironment{table}{\renewcommand{\baselinestretch}{1.0}\selectfont}

\biboptions{sort&compress}

\journal{Computer Physics Communications}
\begin{document}

\begin{frontmatter}

\title{A componentwise spherical Dirichlet-to-Neumann closure for finite-volume MHD at finite magnetic Reynolds number}

\author[1]{Min Ki Jung}
\author[2]{Brian Wynne}
\author[2]{Francisco Saenz}
\author[3]{Yufan Xu}
\author[2,3]{Egemen Kolemen\corref{cor2}}
\cortext[cor2]{ekolemen@princeton.edu, ekolemen@pppl.gov}
\author[1]{Yong-Su Na\corref{cor1}}
\cortext[cor1]{ysna@snu.ac.kr}

\affiliation[1]{organization={Department of Nuclear Engineering, Seoul National University},
            city={Seoul},
            postcode={08826},
            country={South Korea}}
\affiliation[2]{organization={Department of Mechanical and Aerospace Engineering, Princeton University},
            city={Princeton},
            state={New Jersey},
            postcode={08540}, 
            country={USA}}
\affiliation[3]{organization={Princeton Plasma Physics Laboratory},
            city={Princeton},
            state={New Jersey},
            postcode={08540}, 
            country={USA}}

\begin{abstract}
At finite magnetic Reynolds number, the field induced by a moving conducting fluid or solid extends into open space, so a finite mesh needs an outer boundary condition that represents the decaying exterior field. Magnetohydrodynamic (MHD) codes widely use pseudo-vacuum or truncation at a finite radius. Neither condition is physically correct for a vacuum exterior, but the exact alternatives are costly or limited to simple geometries. We present a Dirichlet-to-Neumann (DtN) closure on a fixed enclosing sphere and release it as part of the open-source multi-region finite-volume solver FreeMHD. The closure matches each Cartesian component of the vector potential to its decaying exterior solution, which retains all three decaying vector-Laplace families. A meshed vacuum buffer connects non-spherical conductors to the sphere. We verify the closure with manufactured modes, sphere free decay, sphere and oblate dynamos, and a freely braking rotor. Under refinement, oblate dynamo thresholds and sphere growth rates converge toward independent spectral references, and the rotor torque matches an exact transient reference. On the same sphere, pseudo-vacuum and truncation give the larger errors predicted by their exterior models. The DtN closure achieves the accuracy of a distant outer boundary with a compact domain.
\end{abstract}

\begin{keyword}
Magnetohydrodynamics \sep Finite volume method \sep Vacuum exterior \sep Dirichlet-to-Neumann map \sep Dynamo onset \sep Oblate spheroid
\end{keyword}

\end{frontmatter}

\section{Introduction}\label{sec:introduction}

At finite magnetic Reynolds number $R_m=\mu_0\sigma UL$, defined by the magnetic permeability $\mu_0$, electrical conductivity $\sigma$, and characteristic velocity $U$ and length $L$, motion of a conductor changes its magnetic field on time scales comparable to magnetic diffusion. Although currents close inside the conducting regions, the induced field extends through the surrounding vacuum into open space. A finite-volume calculation therefore needs a condition on its finite outer boundary that represents the decaying exterior solution. A local condition on that boundary, however, behaves as if the conductor were enclosed by a different medium. Its error reaches back into the conductor and alters the induced currents and the Lorentz force, and so changes the coupled evolution of the flow and the magnetic field. Moving the boundary outward reduces this error but enlarges the vacuum mesh and the solve cost. The difficulty is greater in multi-region calculations, in which a moving solid, a fluid, and a vacuum buffer can have separate meshes while their induced fields must remain coupled.

Three treatments of the exterior are common. Boundary-element coupling represents the exterior field on the conductor surface but introduces a dense surface solve and nonlocal communication~\cite{StefaniAN00,XuJCP04,IskakovJCP04,IskakovGAFD05,GieseckeMHD08,BandaruJCP16}. Pseudo-vacuum conditions avoid that solve but force the induced field to be normal to the boundary, as at the surface of an ideal ferromagnet~\cite{HurlburtAJ88,BrandenburgAJ95,RudigerAA01,GuermondEJM03,KenjeresPRL07,GissingerEPL08,HubbardGAFD10,CresswellPRF23}. A meshed vacuum with a simple outer condition accommodates general conductor shapes~\cite{ChanPEPI01,HuangFED02,GailitisPoP04,MatsuiOkudaIJCFD04,GuermondEJM03,KawczynskiPoF18}. Setting the induced vector potential to zero at the outer boundary, for example, makes that boundary act as a perfectly conducting wall, and the resulting truncation error can be reduced only by enlarging the domain. Each choice trades exterior fidelity against geometric freedom, memory, and solver work.

Analytic matching to the exterior solution provides a Dirichlet-to-Neumann (DtN) condition on a sphere, because each decaying scalar harmonic has a known radial derivative at the boundary~\cite{KellerGivoliJCP89,GivoliJCP91}. Spectral dynamo methods apply this relation to the poloidal scalar of the magnetic field~\cite{Elsasser46,BullardGellman54}. Ref.~\citenum{FournierECCOMAS06} implemented a spherical vector-potential DtN map in a Fourier spectral-element solver for axisymmetric domains, whose azimuthal expansion also admits non-axisymmetric fields. In that work the conductor fills the enclosed ball, and the spectral-element solver enforces the Coulomb gauge through the electric potential, with the radial component $A_r$ of the vector potential set to zero on the boundary. The exterior vector potential is then purely toroidal, so the map needs only one of the three families of decaying vector-Laplace solutions. The same work also proposed a meshed buffer around a non-spherical conductor but did not test it. More recently, Fontana et al.~\cite{FontanaCPC22} use a componentwise Robin relation in a Fourier continuation spectral solver for a layer that is periodic in two directions, and enforce the Coulomb gauge in the exterior. Their geometry leads to a condition on Fourier modes rather than to spherical matching.

In this paper we present an exterior closure for an established multi-region finite-volume solver. On a fixed enclosing sphere, the closure matches each Cartesian component of the vector potential to its decaying exterior solution. This componentwise matching retains all three families of decaying vector-Laplace solutions. The spherical map of Ref.~\citenum{FournierECCOMAS06}, the closest prior work, carries only the toroidal family, whereas the present closure also retains the gradient and non-solenoidal families. The gradient family is needed because even a Coulomb-compatible trace generally has a radial component, which only this family can carry. The non-solenoidal family is needed because the discrete trace does not in general satisfy the Coulomb condition. Retaining them keeps the closure consistent with the vector Laplace equation that is actually solved in the vacuum. A meshed buffer connects non-spherical conductors to the sphere, and the resulting boundary derivative is returned to the regional vector-potential solve. Unlike truncation, whose radius must grow as the mesh is refined so that the truncation error stays below the discretization error, the DtN condition is exact on any sphere that encloses the conductors. In our tests the accuracy changes little with the sphere's radius, so the sphere can stay close to the conductor. Because the closure acts on Cartesian components and returns a boundary derivative, it can be added to an existing finite-volume solver as an outer boundary condition. We implement it in the open-source OpenFOAM-based solver \texttt{FreeMHD}~\cite{WynnePoP25}, and the implementation is released with the code~\cite{FreeMHDcode}.

This work builds on Jung et al.~\cite{JungPoP26}, who extended \texttt{FreeMHD} beyond the inductionless approximation by solving the induction equation in a vector potential formulation, and verified and validated the extended solver. They closed the exterior with zero induced vector potential on a surrounding vacuum region and deferred a more general treatment to a separate publication. The present paper provides that treatment and is concerned only with the exterior boundary. We test the exterior update against manufactured modes, sphere free decay, sphere dynamo growth, oblate dynamo onset, and magnetic braking of a rigid rotor. Free surfaces may lie inside the enclosing sphere, but no free-surface response is tested here.

The paper is organized as follows. Section~\ref{sec:problem} states the regional problem, and Section~\ref{sec:closure} constructs the boundary map. Section~\ref{sec:verification} gives the verification strategy and the common settings and then presents the five tests. Section~\ref{sec:cost} weighs accuracy against cost in oblate onset and the braking rotor. Section~\ref{sec:limits} discusses the limitations, and Section~\ref{sec:conclusions} concludes.

\section{Problem and computational domain}\label{sec:problem}

The computational domain consists of conducting regions, which may be moving fluids or solids, and vacuum regions, by which we mean any electrically insulating region. In each conductor we solve for the induced vector potential $\mathbf A\equiv\mathbf A_{\mathrm{ind}}$ and the electric potential $\phi$, whereas in vacuum regions we solve for $\mathbf A$ only. The magnetic field is $\mathbf B=\nabla\times\mathbf A+\mathbf B_{\mathrm{ext}}$, where $\mathbf B_{\mathrm{ext}}$ is prescribed and curl-free. With electrical conductivity $\sigma$, magnetic permeability $\mu_0$, velocity $\mathbf u$, and current density $\mathbf j$, the conductor equations relevant to the exterior coupling are
\begin{align}
\partial_t\mathbf A &=-\nabla\phi+\mathbf u\times\mathbf B+(\mu_0\sigma)^{-1}\nabla^2\mathbf A,\label{eq:induction}\\
\mathbf j&=\sigma(-\partial_t\mathbf A-\nabla\phi+\mathbf u\times\mathbf B),\label{eq:ohm}\\
\nabla\cdot(\sigma\nabla\phi)&=\nabla\cdot[\sigma(\mathbf u\times\mathbf B-\partial_t\mathbf A)].\label{eq:potential}
\end{align}
The potential equation ensures $\nabla\cdot\mathbf j=0$. The full fluid equations, the regional discretization, and the pressure-velocity coupling are given by Jung et al.~\cite{JungPoP26}.

In a vacuum region the implemented magnetic equation is
\begin{equation}\label{eq:vacuum}
\nabla^2\mathbf A=0.
\end{equation}
Where $\sigma$ is uniform, as in every region tested here, the continuum equations imply $\nabla^2(\nabla\cdot\mathbf A)=0$, so the Coulomb gauge $\nabla\cdot\mathbf A=0$ holds when compatible interface and boundary conditions make this harmonic divergence vanish. The discrete calculation does not impose the gauge.

At a conductor-vacuum interface with unit normal vector $\mathbf n$, the normal component of the conductor-side Ohmic current, Eq.~\eqref{eq:ohm}, vanishes,
\begin{equation}\label{eq:insulating}
\mathbf n\cdot\mathbf j=\sigma\mathbf n\cdot(-\partial_t\mathbf A-\nabla\phi+\mathbf u\times\mathbf B)=0.
\end{equation}
The motional term remains where a prescribed tangential flow reaches the wall. Across conducting interfaces, by contrast, the normal Ohmic current is continuous. Regional coupling transfers $\mathbf A$ and its normal derivative through mapped patches~\cite{JungPoP26}. For equal permeability and no surface current, these transfers match the continuum interface conditions
\begin{equation}\label{eq:interfaces}
[\![\mathbf A]\!]=0,\qquad [\![\partial_n\mathbf A]\!]=0,\qquad
[\![\mathbf n\cdot\mathbf j]\!]=0
\quad\text{across conducting interfaces}.
\end{equation}
Here $[\![\cdot]\!]$ is the jump across the interface and $\partial_n\equiv\mathbf n\cdot\nabla$, both taken along a single common normal. The vacuum side of Eq.~\eqref{eq:insulating} carries no Ohmic current.

The solver couples $\phi$ and $\mathbf A$ by a partitioned iteration with $n_c$ outer correctors per time step~\cite{JungPoP26}. At the first corrector, the start-of-step predictor, when enabled, supplies the scalar-potential source and the insulating condition with the final $\partial_t\mathbf A$ of the preceding step. Without it, they use the Euler difference, which is zero at that corrector because $\mathbf A$ has not yet changed. Later correctors use the latest Euler difference. A conductor is solved either as a fluid region or as a solid region of \texttt{FreeMHD}~\cite{JungPoP26}, and we call the two the velocity-bearing and solid induction paths. They differ in how $\partial_t\mathbf A$ enters the potential equation, Eq.~\eqref{eq:potential}. The velocity-bearing path carries the velocity $\mathbf u$, so every moving conductor uses it, and it keeps the volume source $-\nabla\cdot(\sigma\partial_t\mathbf A)$. The solid path has $\mathbf u=0$ and drops this source, which vanishes in the continuum when $\sigma$ is uniform and $\nabla\cdot\mathbf A=0$. In the solid path, $\partial_t\mathbf A$ therefore enters the potential solve only through the Neumann value of $\phi$ that imposes the insulating condition, Eq.~\eqref{eq:insulating}, on the conductor surface, and this value is updated at each corrector from the latest $\partial_t\mathbf A$.

\begin{figure}[!htbp]
\centering
\includegraphics[width=\linewidth]{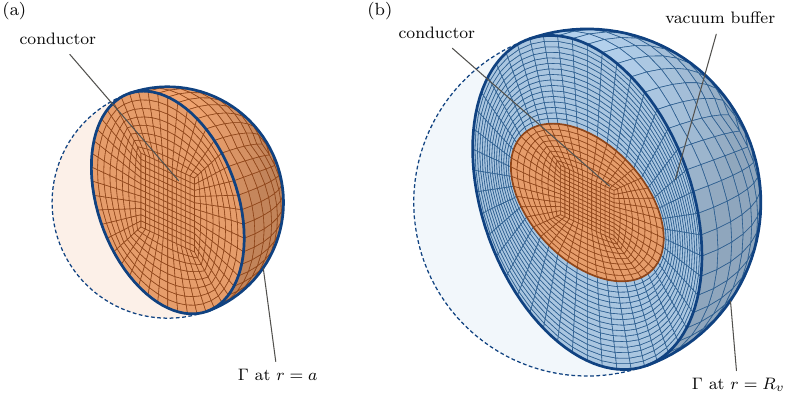}
\caption{Regional arrangements, cut open to show the mesh. (a) Conducting sphere of radius $a$ with $\Gamma$ on its surface, so that $R_v=a$ and no vacuum is meshed. (b) Oblate conductor with equatorial and polar semi-axes $a$ and $c=a/1.5$, inside a meshed vacuum buffer that extends to $\Gamma$ at $R_v=1.5a$. Dashed curves outline the removed halves. Meshes are shown at half the resolution of $N24$.}
\label{fig:geometry}
\end{figure}

All current-carrying regions lie inside a sphere $\Gamma$ of radius $R_v$, the vacuum radius, outside which the vacuum is left unmeshed. The outer condition is imposed on $\Gamma$, by the DtN map or, for comparison, by a local condition such as pseudo-vacuum or $\mathbf A_{\rm ind}=0$. If the conductor surface is itself spherical, the DtN operator acts on that surface and no vacuum mesh is needed. A non-spherical conductor cannot supply the spherical trace directly, so it is enclosed by a meshed vacuum buffer that extends to $\Gamma$. The buffer solves Eq.~\eqref{eq:vacuum}, and the operator continues the buffer's boundary trace into the unmeshed exterior, where the continuation satisfies $\mathbf A\to0$ as the distance $r$ from the center of $\Gamma$ tends to infinity. The operator acts on $\mathbf A$ alone and prescribes neither $\phi$ nor a magnetic scalar potential. Figure~\ref{fig:geometry} shows both arrangements. The numerical boundary is a polyhedral approximation of the sphere, as specified in Section~\ref{sec:closure}.

\section{Spherical DtN closure}\label{sec:closure}

The closure replaces the unmeshed exterior by a Neumann condition on $\Gamma$. Given the trace of $\mathbf A$ on $\Gamma$, it returns the normal derivative of the decaying exterior solution of Eq.~\eqref{eq:vacuum} that has this trace. Section~\ref{sec:procedure} states the equations evaluated at each boundary update and their sequence. Section~\ref{sec:exactness} shows that the map is exact for all three families of decaying vector-Laplace solutions and explains why this exactness alone does not guarantee a current-free exterior. Section~\ref{sec:implementation} gives the discretization details and cost.

\subsection{Boundary update}\label{sec:procedure}

We write $R$ for the radius of $\Gamma$, so that $R=R_v$, and use spherical coordinates $(r,\theta,\varphi)$ centered on $\Gamma$, with polar angle $\theta$ and azimuth $\varphi$. Each Cartesian component of a decaying solution of Eq.~\eqref{eq:vacuum} is a scalar harmonic function outside $\Gamma$, so it is fixed by its trace and expands in decaying solid harmonics. Let $Y_{lm}(\theta,\varphi)$ be the orthonormal complex scalar spherical harmonics of degree $l$ and order $m$, and let $A_i^b$ be the trace of the Cartesian component $i\in\{x,y,z\}$ on $\Gamma$. Then
\begin{equation}\label{eq:projection}
 s_{i,lm}=\int_{\Gamma}A_i^bY_{lm}^{*}\,d\Omega,
 \qquad A_i(r,\theta,\varphi)=\sum_{l=0}^{L_{\mathrm{SH}}}\sum_{m=-l}^{l}s_{i,lm}(R/r)^{l+1}Y_{lm}(\theta,\varphi),
\end{equation}
where $d\Omega=\sin\theta\,d\theta\,d\varphi$ is the solid-angle element and the truncation degree $L_{\mathrm{SH}}$ sets the angular resolution of the map. Each term $(R/r)^{l+1}Y_{lm}$ is a decaying solid harmonic of degree $l$, that is, a solution of Laplace's equation outside $\Gamma$ that vanishes at infinity. Differentiating at $r=R$ gives the scalar DtN map
\begin{equation}\label{eq:radial-map}
 \left.\partial_r A_i\right|_{r=R}= -\sum_{l=0}^{L_{\mathrm{SH}}}\sum_{m=-l}^{l}\frac{l+1}{R}s_{i,lm}Y_{lm},\qquad i\in\{x,y,z\}.
\end{equation}
Equation~\eqref{eq:projection} automatically retains all three families of decaying vector solutions without decomposing $\mathbf A$, and Section~\ref{sec:exactness} proves that the map is exact for each of them. A closure that neglects one of them, for instance the non-solenoidal family, would instead have to decompose the trace into the individual vector families. The same expansion supplies the two tangential derivatives,
\begin{equation}\label{eq:tangential}
 \left.\partial_\theta A_i\right|_{r=R}=\sum_{l=0}^{L_{\mathrm{SH}}}\sum_{m=-l}^{l}s_{i,lm}\,\partial_\theta Y_{lm},
 \qquad
 \left.\partial_\varphi A_i\right|_{r=R}=\sum_{l=0}^{L_{\mathrm{SH}}}\sum_{m=-l}^{l}s_{i,lm}\,\partial_\varphi Y_{lm}.
\end{equation}
A finite-volume face normal $\mathbf n$ on $\Gamma$ can deviate from the radial unit vector $\hat{\mathbf r}$, so the Neumann datum imposed on each face is the full face-normal derivative
\begin{equation}\label{eq:face-normal}
 \mathbf g^{\mathrm{DtN}}=\mathbf n\cdot\nabla\mathbf A
 =n_r\partial_r\mathbf A+\frac{n_\theta}{R}\partial_\theta\mathbf A
 +\frac{n_\varphi}{R\sin\theta}\partial_\varphi\mathbf A,
\end{equation}
where $n_r$, $n_\theta$, and $n_\varphi$ are the spherical components of $\mathbf n$. On the mesh, the integral in Eq.~\eqref{eq:projection} becomes the face sum
\begin{equation}\label{eq:face-projection}
s_{i,lm}\simeq\sum_{f\subset\Gamma}A^b_{i,f}Y_{lm}^{*}(\theta_f,\varphi_f)\frac{w_f}{R_f^2},
\end{equation}
where $A^b_{i,f}$ is the trace at the centroid of face $f$, $(\theta_f,\varphi_f)$ are the centroid angles, $w_f$ is the face area, and $R_f$ is the centroid radius.

Algorithm~\ref{alg:boundary} combines these steps into an update of the Neumann data on $\Gamma$. The update runs inside the solve of whichever region carries $\Gamma$, the conductor or the vacuum buffer. Each pass projects the current trace with Eq.~\eqref{eq:face-projection} and evaluates Eqs.~\eqref{eq:radial-map}, \eqref{eq:tangential}, and \eqref{eq:face-normal} on every face of $\Gamma$, which gives $\mathbf g_f^{\mathrm{DtN}}$ on face $f$. Instead of imposing this value directly, the pass relaxes the face-normal gradient $\mathbf g_f$ imposed on the face toward it,
\begin{equation}\label{eq:relaxation}
\mathbf g_f\gets(1-\gamma)\,\mathbf g_f+\gamma\,\mathbf g_f^{\mathrm{DtN}},
\end{equation}
where $0<\gamma\le1$ is the relaxation factor. The relaxation acts on the imposed derivative, not on the harmonic coefficients, and a fixed point of Eq.~\eqref{eq:relaxation} satisfies $\mathbf g_f=\mathbf g_f^{\mathrm{DtN}}$ for any $\gamma$. Passes repeat until the residual of the regional solve meets its tolerance or an iteration limit is reached. Only the $3(L_{\mathrm{SH}}+1)^2$ scalar coefficients $s_{i,lm}$ are computed, and no vector spherical harmonic coefficient is formed.

\begin{algorithm}[!htbp]
\caption{Boundary update on $\Gamma$ within a regional vector-potential solve}
\label{alg:boundary}
\renewcommand{\baselinestretch}{1.0}\small
\begin{algorithmic}[1]
\State At first use, tabulate $Y_{lm}$ for $l\le L_{\mathrm{SH}}$ at each local face centroid and tangential-difference point.
\Repeat
\State Solve the equation for $\mathbf A$ in the region whose outer surface is $\Gamma$, with face-normal gradient $\mathbf g_f$ on $\Gamma$.
\State Compute $s_{i,lm}$ for $i\in\{x,y,z\}$ from the face sum of Eq.~\eqref{eq:face-projection} and sum them over processes.
\State On each face, evaluate $\partial_r\mathbf A$ from Eq.~\eqref{eq:radial-map} and the tangential derivatives from Eq.~\eqref{eq:tangential}, using the tabulated harmonics.
\State Form $\mathbf g_f^{\mathrm{DtN}}$ from Eq.~\eqref{eq:face-normal}.
\State Set $\mathbf g_f\gets(1-\gamma)\mathbf g_f+\gamma\mathbf g_f^{\mathrm{DtN}}$.
\Until{the regional solve residual meets the iteration tolerance or the iteration limit is reached}
\end{algorithmic}
\end{algorithm}

The expansion can also be inverted to give a Neumann-to-Dirichlet (NtD) closure. Because the factor $-(l+1)/R$ in Eq.~\eqref{eq:radial-map} is nonzero for every $l\geq0$, projecting a prescribed $\partial_r A_i$ onto $Y_{lm}$ and multiplying each coefficient by $-R/(l+1)$ recovers $s_{i,lm}$. Equation~\eqref{eq:projection} then gives the trace to impose as Dirichlet data on $\Gamma$. On a polyhedral $\Gamma$, the tangential terms of Eq.~\eqref{eq:face-normal} depend on the trace itself, so an NtD update would need to lag or iterate them. This variant is not tested here.

\subsection{Exactness for all three vector families}\label{sec:exactness}

The map treats $\mathbf A$ as three scalars. To show which vector fields it represents, we expand the same exterior solution in vector spherical harmonics (VSH). The coefficients introduced here serve only this analysis, and the closure never computes them. Define $\mathbf Y_{lm}=Y_{lm}\hat{\mathbf r}$, $\boldsymbol\Psi_{lm}=r\nabla Y_{lm}$, and $\boldsymbol\Phi_{lm}=\hat{\mathbf r}\times\boldsymbol\Psi_{lm}$. Every decaying solution of Eq.~\eqref{eq:vacuum} outside $\Gamma$ can be written as
\begin{equation}\label{eq:families}
\begin{split}
\mathbf A=\sum_{l,m}\bigg[&a_{lm}(R/r)^{l+1}\boldsymbol\Phi_{lm}\\
&+b_{lm}(R/r)^{l+2}\{\boldsymbol\Psi_{lm}-(l+1)\mathbf Y_{lm}\}\\
&+c_{lm}(R/r)^l\{\boldsymbol\Psi_{lm}+l\mathbf Y_{lm}\}\bigg].
\end{split}
\end{equation}
We call the $a_{lm}$, $b_{lm}$, and $c_{lm}$ terms the toroidal, gradient, and non-solenoidal families, respectively. The toroidal and non-solenoidal families start at $l=1$, whereas the gradient family also includes $l=0$. The toroidal family carries the poloidal magnetic field of a current-free continuum exterior. The gradient family equals $R\nabla[(R/r)^{l+1}Y_{lm}]$, so it changes $\mathbf A$ without carrying a magnetic field. The non-solenoidal family is the only one with nonzero divergence, and later in this section, we explain why the map retains it. 

The trace fixes all three amplitudes. Only the toroidal family has a component along $\boldsymbol\Phi_{lm}$, so its amplitude follows directly from the trace,
\begin{equation}\label{eq:tor-coeff}
a_{lm}=\frac{1}{l(l+1)}\int_{\Gamma}\mathbf A^b\cdot\boldsymbol\Phi_{lm}^{*}\,d\Omega,\qquad l\geq1,
\end{equation}
where $\mathbf A^b$ is the trace on $\Gamma$. The gradient and non-solenoidal families share the directions $\mathbf Y_{lm}$ and $\boldsymbol\Psi_{lm}$, along which the trace has the components
\begin{equation}\label{eq:trace-coeffs}
\alpha_{lm}=\int_{\Gamma}\hat{\mathbf r}\cdot\mathbf A^b\,Y_{lm}^{*}\,d\Omega,\qquad
\beta_{lm}=\frac{1}{l(l+1)}\int_{\Gamma}\mathbf A^b\cdot\boldsymbol\Psi_{lm}^{*}\,d\Omega,
\end{equation}
with $\beta_{lm}$ defined for $l\geq1$. Evaluating Eq.~\eqref{eq:families} at $r=R$ gives $\alpha_{lm}=-(l+1)b_{lm}+lc_{lm}$ and $\beta_{lm}=b_{lm}+c_{lm}$, so for $l\geq1$
\begin{align}
 b_{lm}&=\frac{l\beta_{lm}-\alpha_{lm}}{2l+1},\label{eq:grad-coeff}\\
 c_{lm}&=\frac{\alpha_{lm}+(l+1)\beta_{lm}}{2l+1},\label{eq:third-coeff}
\end{align}
and $b_{00}=-\alpha_{00}$ at $l=0$. A Coulomb-compatible trace has no non-solenoidal content, that is,
\begin{equation}\label{eq:compatibility}
 \alpha_{lm}+(l+1)\beta_{lm}=0,\qquad l\geq1.
\end{equation}
The compatibility residual $\alpha_{lm}+(l+1)\beta_{lm}$ is thus exactly $(2l+1)c_{lm}$, so the non-solenoidal family carries precisely the part of the trace that violates Eq.~\eqref{eq:compatibility}. 

The radial powers in Eq.~\eqref{eq:families} give the exact normal derivatives at $\Gamma$, namely $-(l+1)/R$, $-(l+2)/R$, and $-l/R$ for the toroidal, gradient, and non-solenoidal families. The three families of the same degree $l$ therefore need three different factors. A map that multiplies all VSH components of degree $l$ by one factor, such as $-(l+1)/R$, is exact for the toroidal family but not for the other two. The componentwise map is exact because the Cartesian components of each family term are decaying solid harmonics of a single degree, namely $l$, $l+1$, and $l-1$ for the toroidal, gradient, and non-solenoidal families, respectively. Equation~\eqref{eq:radial-map} applies the correct factor to each of these degrees. For example, the $l=1$ non-solenoidal member with $m=0$ is $(R/r)\hat{\mathbf z}$, whose Cartesian components are solid harmonics of degree zero. For this reason the sums in Eq.~\eqref{eq:projection} start at $l=0$, and $L_{\mathrm{SH}}$ must reach degree $l+1$ to represent a gradient mode of degree $l$. The Cartesian projection of Eq.~\eqref{eq:projection} also needs no angular derivatives of $Y_{lm}$, whereas the projections in Eq.~\eqref{eq:trace-coeffs} would.

The non-solenoidal family deserves separate comment, because a physical vacuum does not contain it. The family has nonzero curl and carries a toroidal magnetic field, so in the unconstrained vector-Laplace problem it can produce a formal current $\mu_0^{-1}\nabla\times\mathbf B$. A physical continuum vacuum is current-free, and in the Coulomb gauge its vector potential contains only the toroidal and gradient families. The discrete solution, however, does not enforce the Coulomb gauge, so a discrete trace generally has a nonzero compatibility residual and hence non-solenoidal content. A map built only from the toroidal and gradient families assigns no derivative to this content. This includes the $l=1$ member, which decays as $1/r$, more slowly than the corresponding toroidal dipole. Retaining the non-solenoidal family gives this part of the trace its own decaying continuation and hence the correct normal derivative, whereas discarding it would change the derivative applied to a solution of Eq.~\eqref{eq:vacuum}. We therefore retain all three decaying families admitted by the implemented vector Laplace equation, and Section~\ref{sec:mms} tests this distinction directly. 

\subsection{Discretization and cost}\label{sec:implementation}

The weight $w_f/R_f^2$ in Eq.~\eqref{eq:face-projection} omits the factor $\mathbf n\cdot\hat{\mathbf r}$ of the solid-angle element. For shape-regular faces of diameter $h_f$ with vertices on the sphere, the normal deviates from $\hat{\mathbf r}$ by $O(h_f/R_f)$, so the omission causes a relative weight error of $O(h_f^2/R_f^2)$. Each face contributes directly, without interpolation onto a separate spherical grid, and the reconstruction uses the local $R_f$. The derivatives in Eq.~\eqref{eq:tangential} are evaluated by centered differences of the reconstructed expansion, with a step of $10^{-6}$ radian in each angular direction. The azimuthal term of Eq.~\eqref{eq:face-normal} retains its $1/\sin\theta$ factor. For a real trace, negative orders satisfy $s_{i,l,-m}=(-1)^m s_{i,lm}^{*}$, but the implementation still sums all orders $-l\le m\le l$ and keeps both the real and imaginary parts of the coefficients.

With $N_b$ boundary faces and $M=(L_{\mathrm{SH}}+1)^2$ scalar modes, projection and reconstruction cost $O(N_bM)=O(N_bL_{\mathrm{SH}}^2)$ per application. Because the geometry is fixed, the harmonics are tabulated once and reused. Each process stores them only for its local boundary faces, and the only global communication is a reduction of $3M$ complex coefficients, independent of the number of volume cells. The cost of the boundary operation is thus separate from that of the regional linear solves, which depends on the conductor and buffer meshes. The finite degree $L_{\mathrm{SH}}$ truncates the trace, so omitted scalar degrees receive no reconstructed derivative. Because the degree-$l$ content of each Cartesian component decays as $r^{-(l+1)}$ outside any sphere that shares the center of $\Gamma$ and encloses the conductors, a sphere $\Gamma$ placed farther from them carries less high-degree content and needs a lower $L_{\mathrm{SH}}$. Since the boundary cost grows only as $(L_{\mathrm{SH}}+1)^2$ per face, $L_{\mathrm{SH}}$ can simply be raised until the quantity of interest no longer changes, provided it stays well below the angular resolution of the faces on $\Gamma$. Section~\ref{sec:mms} gives the degree that each family needs, and Section~\ref{sec:dynamo} applies this check to the value used in the coupled tests.

\section{Verification}\label{sec:verification}

We verify the closure in five tests of increasing scope, each against an analytic or independently computed reference. Each test adds one ingredient to the previous one, so that a discrepancy can be traced to that ingredient. Manufactured modes test the map alone, for all three vector families, without time stepping (Section~\ref{sec:mms}). Sphere free decay places the map inside the time-dependent induction solve, with a meshed vacuum buffer and no flow, and compares with the analytic decay time (Section~\ref{sec:decay}). The sphere dynamo adds a prescribed flow and applies the map directly on the conductor surface, where the trace is not Coulomb compatible, and compares growth rates and the onset threshold with an independent spectral solution (Section~\ref{sec:dynamo}). Oblate onset replaces the sphere by a spheroid inside a meshed buffer, the configuration the closure is designed for, and compares with published thresholds (Section~\ref{sec:onset}). Finally, the braking rotor couples the field to the rotation of a rigid sphere and compares the transient torque and speed with an exact solution (Section~\ref{sec:rotor}). Along the way, the decay, dynamo, oblate, and rotor tests compare DtN with close truncation, which imposes $\mathbf A_{\rm ind}=0$ on the same outer boundary. The dynamo and rotor tests also compare it with pseudo-vacuum, and the rotor test with a graded shell. Section~\ref{sec:cost} then weighs accuracy against cost in the oblate and rotor tests.

\subsection{Common settings and error measures}\label{sec:protocol}

In the coupled tests the conductor is a sphere of radius $a$, except in oblate onset, where it is an oblate spheroid of equatorial semi-axis $a$. Lengths are in units of $a$ and times in units of the magnetic diffusion time $\mu_0\sigma a^2$. Table~\ref{tab:settings} lists the settings of each test. The coupled tests use three block-structured hexahedral meshes of the conductor with the same topology. A grid in spherical coordinates degenerates at the center and on the polar axis, so each conductor mesh is an O-grid. A cube of $N\times N\times N$ cells fills the core, and six curved blocks connect its faces to the conductor surface, each with $N\times N$ cells there. This surface therefore carries $6N^2$ faces. For oblate onset, the same meshes are mapped onto the spheroid. The meshes $N16$, $N24$, and $N36$ have $N=16$, $24$, and $36$, with $1536$, $3456$, and $7776$ surface faces and $11$, $16$, and $24$ radial cells in the six curved blocks. Each refinement step thus increases the resolution by a factor of about $1.5$ in every direction. The $N24$ mesh is the production mesh, and the other two serve for refinement studies.

For each test, Table~\ref{tab:settings} also gives the number $n_c$ of outer correctors, the induction path (Section~\ref{sec:problem}), and the gradient scheme used to derive the magnetic field and current. The boundary update of Algorithm~\ref{alg:boundary} is applied inside each regional vector-potential solve, with $\gamma=0.1$ in all reported calculations. In these runs the residual of the regional solve is already below the tolerance of $10^{-3}$ on entry, so each call makes a single pass. With a buffer, the update is thus applied once per outer corrector, and on a conductor surface once per inner vector-potential iteration. The outer correctors therefore also iterate the boundary update. The convergence criterion of the passes, however, is the residual of the regional solve, which does not measure how closely $\mathbf g_f$ matches $\mathbf g_f^{\mathrm{DtN}}$. We therefore also monitored the DtN mismatch, the relative surface $L^2$ difference between the imposed $\mathbf g_f$ and a freshly reconstructed $\mathbf g_f^{\mathrm{DtN}}$. Sampled at the end of every 100th time step, the mismatch stays below $3.2\times10^{-5}$ in the monitored free-decay and oblate runs. In the sphere dynamo, sampled at the end of every 500th time step, it stays below $1.3\times10^{-4}$.

Let $E_B=\frac12\int_{\mathrm{conductor}}|\nabla\times\mathbf A|^2\,dV$ measure the magnetic energy in the conductor. For a dynamo, the instantaneous growth rate after the initial transient is $\lambda(t)=\frac12\,d\ln E_B/dt$. Where it approaches a limit $\lambda_\infty$, we fit $\lambda(t)=\lambda_\infty-Ce^{-\kappa t}$ with fit constants $C$ and $\kappa$ and report $\lambda_\infty$. The growth-rate fit uses an approach limit only when that limit stays stable as the start time of the fit is moved, and otherwise uses a late-time energy slope. The onset threshold $R_c$ is the zero of a quadratic through the growth rates at three values of $R_m$ that bracket it. Its relative difference from a reference threshold $R_c^{\mathrm{ref}}$ is
\begin{equation}\label{eq:reldiff}
\left(\frac{R_c}{R_c^{\mathrm{ref}}}-1\right)\times100\%.
\end{equation}
With this definition the sign remains meaningful when DtN is compared with a shell closed by $\mathbf A_{\rm ind}=0$.

For a refinement series with mesh spacing $h\propto N^{-1}$ and refinement ratio $1.5$, we compute the observed order from three mesh values $Q_1$, $Q_2$, and $Q_3$, ordered from coarse to fine, as $p=\log[(Q_1-Q_2)/(Q_2-Q_3)]/\log1.5$. If the discretization error is proportional to $h^p$, Richardson extrapolation gives the mesh-converged estimate, or Richardson value, $Q_{\mathrm R}=Q_3+(Q_3-Q_2)/(1.5^p-1)$. We evaluate $Q_{\mathrm R}$ both at the observed order and at $p=2$, the formal order of the spatial discretization. The two values serve as consistency estimates, never as error bounds. No temporal order is inferred from the spatial refinement series.

\begin{table}[!htbp]
\centering
\caption{Settings of the verification tests. Cell counts are for the conductor and selected vacuum meshes. Radii are in units of $a$, the sphere radius or the equatorial semi-axis of the spheroid, whose polar semi-axis is $c$. Graded shells have geometrically growing radial spacing up to $R_v$. LS and Gauss denote unlimited least-squares and Gauss-linear gradients. All tests use $L_{\rm SH}=8$. The start-of-step predictor is used for off-threshold growth rates and the rotor, but not for onset thresholds or the comparison of outer conditions in Section~\ref{sec:dynamo}.}
\label{tab:settings}
\scriptsize
\setlength{\tabcolsep}{2.5pt}
\begin{tabular}{@{}>{\raggedright\arraybackslash}p{0.15\linewidth}>{\raggedright\arraybackslash}p{0.36\linewidth}>{\raggedright\arraybackslash}p{0.14\linewidth}>{\raggedright\arraybackslash}p{0.07\linewidth}>{\raggedright\arraybackslash}p{0.09\linewidth}>{\raggedright\arraybackslash}p{0.08\linewidth}@{}}
\hline
Case & mesh and cells or boundary faces & exterior & $n_c$ & $\Delta t$ & grad., path \\
\hline
Manufactured modes & $3456$, $13\,824$ faces & $R_v=2$ & n.a. & n.a. & n.a. \\
Sphere free decay & $N16/24/36$ conductor: $20\,992/69\,120/233\,280$; $N24$, $R_v=1.5$: $69\,120$ shell & $R_v=1.5$ to $3$ & $24$ & $2\times10^{-4}$ & Gauss, solid \\
Sphere dynamo & $N16/24/36$: $20\,992/69\,120/233\,280$ & $R_v=1$; shell check $1.5$ & $20$ & $2\times10^{-4}$ & LS, velocity \\
Oblate onset & $N16/24/36$: $20\,992/69\,120/233\,280$; $N24$ shells: $93\,312$ ($a/c=1.2$), $114\,048$ ($1.5$) & $R_v=1.5$; check $2$ & $20$ & $2\times10^{-4}$ & LS, velocity \\
Braking rotor & $N16/24/36$ conductor: $20\,992/69\,120/233\,280$; $N24$ shell $69\,120$; graded $176\,256$ & $R_v=1.5,2$; graded $R_v=15$ & $20$ & $5\times10^{-4}$ & LS, velocity \\
\hline
\end{tabular}
\end{table}

\subsection{Manufactured modes}\label{sec:mms}

Applied in isolation, the map recovers all three decaying families down to the error floor of the surface quadrature. We prescribe a single mode of Eq.~\eqref{eq:families}, formed from $\operatorname{Re}Y_{lm}$, on the boundary $\Gamma$ of a shell of radius $R=2$, apply the map once, and compare the imposed radial derivative with the analytic one. The angular derivatives of the prescribed trace are formed by central differences, independently of the operator routines. The relative surface error is $\varepsilon=\|\partial_r\mathbf A_{\rm num}-\partial_r\mathbf A_{\rm exact}\|_{L^2(\Gamma)}/\|\partial_r\mathbf A_{\rm exact}\|_{L^2(\Gamma)}$.

\begin{figure}[!htbp]
\centering
\includegraphics[width=\linewidth]{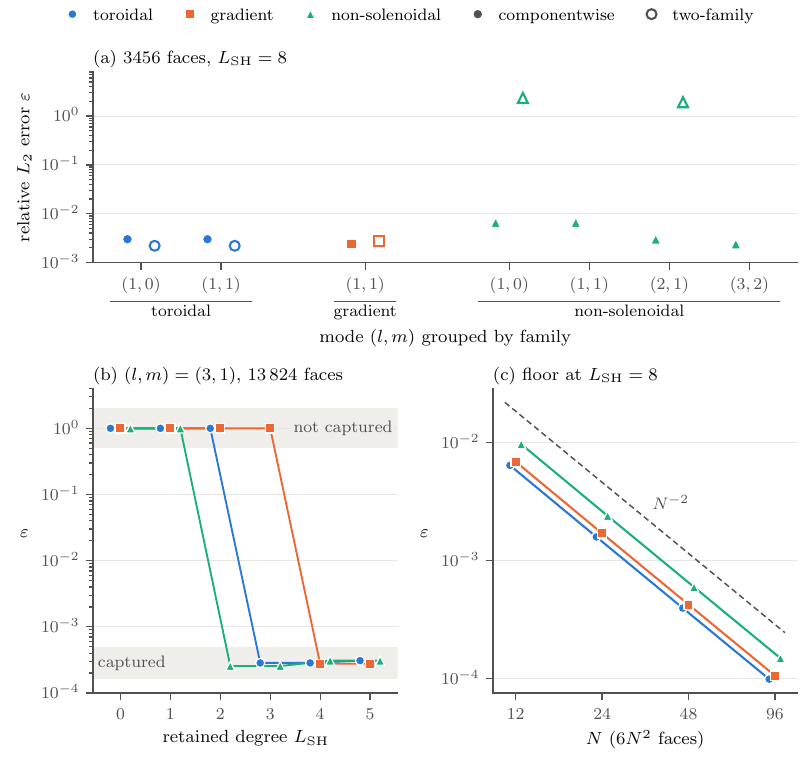}
\caption{Manufactured derivative error $\varepsilon$ for single modes on $\Gamma$ at $R=2$, with $L_{\rm SH}=8$ unless varied. (a) Componentwise (filled) and two-family (open) maps on $3456$ faces. No two-family values exist for the non-solenoidal $(1,1)$ and $(3,2)$ modes. (b) Error against $L_{\rm SH}$ for $(l,m)=(3,1)$ in each family on $13\,824$ faces. (c) Error floor of the captured modes against $N$, with $6N^2$ faces and an $N^{-2}$ guide. Markers in (b) and (c) are offset horizontally for clarity.}
\label{fig:operator}
\end{figure}

For axisymmetric and non-axisymmetric modes up to $l=3$, the error is $\varepsilon=0.24\%$ to $0.66\%$ on the boundary with $3456$ faces (Fig.~\ref{fig:operator}(a), filled markers), and each halving of the face spacing reduces it by a factor of $4.00$ to $4.01$.

The truncation degree $L_{\rm SH}$ limits which modes the map captures. A mode that the expansion does not resolve gives $\varepsilon=1$. For $(l,m)=(3,1)$, the non-solenoidal, toroidal, and gradient families enter the Cartesian expansion at $L_{\rm SH}=2$, $3$, and $4$, respectively (Fig.~\ref{fig:operator}(b)). At $L_{\rm SH}=8$, the error floor of the captured modes falls by a factor of $4.00$ to $4.05$ per halving of the face spacing, from $864$ to $55\,296$ faces (Fig.~\ref{fig:operator}(c)).

The non-solenoidal family is required to impose the correct derivative on a general discrete trace. The two-family map gives errors of $0.22\%$ to $0.27\%$ for the tested toroidal and gradient modes but $192\%$ to $235\%$ for non-solenoidal modes, as Fig.~\ref{fig:operator}(a) shows. Single modes cannot bound the harmonic degree that a coupled trace requires, so the coupled sphere checks in Section~\ref{sec:dynamo} test that choice.

\subsection{Sphere free decay}\label{sec:decay}

Free decay of a dipole field in a sphere of radius $a$ tests the closure with a meshed buffer against a known decay time, using the solid induction path. We initialize the fundamental poloidal mode, whose analytic decay time is $\tau_{\mathrm{ref}}=\mu_0\sigma a^2/\pi^2$~\cite{Moffatt78}, or $1/\pi^2$ in our units. In the production case the map closes a shell at $R_v=1.5$ (Table~\ref{tab:settings}). A straight-line fit to $\ln E_B(t)$ over $0.05\leq t\leq0.3$ gives the fitted decay time $\tau_{\mathrm{fit}}=-2/(\mathrm{slope})$, and we report the signed error $(\tau_{\mathrm{fit}}/\tau_{\mathrm{ref}}-1)\times100\%$.

\begin{figure}[!htbp]
\centering
\includegraphics[width=0.65\linewidth]{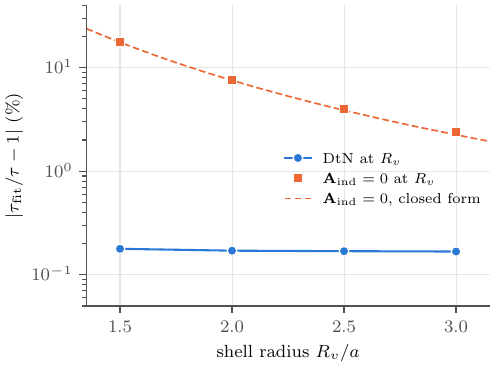}
\caption{Sphere free decay on $N24$ with $n_c=24$. Magnitude of the fitted decay-time error against $R_v$ for DtN (circles) and $\mathbf A_{\rm ind}=0$ (squares), with fits over $0.05\leq t\leq0.3$. The dashed curve is the predicted error for $\mathbf A_{\rm ind}=0$ from Eq.~\eqref{eq:trunc-eigenvalue}. All signed errors are negative.}
\label{fig:decaytrunc}
\end{figure}

The production case gives an error of $-0.178\%$ relative to the analytic decay time (Fig.~\ref{fig:decaytrunc}, $R_v=1.5$), and its DtN mismatch at the end of every 100th step (Section~\ref{sec:protocol}) stays below $3.2\times10^{-5}$. Three-mesh comparisons at fixed time step $\Delta t=2\times10^{-4}$ and $10^{-4}$, fitted over $0.05\leq t\leq0.15$, give observed spatial orders of $1.54$ and $1.49$ against the same reference. These data do not establish a temporal order.

In the continuum, the term $-\nabla\cdot(\sigma\partial_t\mathbf A)$ that the solid path omits (Section~\ref{sec:problem}) vanishes for this axial mode. On the production mesh, the velocity-bearing path, which retains the term, gives $-0.1741\%$ compared with $-0.1781\%$ for the solid path. This benchmark therefore verifies the exterior closure on the solid path, and its refinement orders describe that path. 

Setting $\mathbf A_{\rm ind}=0$ at the shell radius instead gives decay-time errors about $99$ times larger than DtN at $R_v=1.5$ and $14$ times larger at $3$ (Fig.~\ref{fig:decaytrunc}). At $R_v=1.5$, $2$, $2.5$, and $3$ these errors are $-17.57\%$, $-7.62\%$, $-4.00\%$, and $-2.39\%$, whereas the DtN errors span only $0.010$ percentage points across the same radii. We do not interpret this small DtN trend with radius, because the tangential size of the outer cells grows with radius at fixed angular resolution.

Setting $\mathbf A_{\rm ind}=0$ moves the radial wavenumber $k$ of the $l=1$ decay mode from $\pi$ to the first root of
\begin{equation}\label{eq:trunc-eigenvalue}
\frac{k^2\sin k}{\sin k-k\cos k}=-\frac{3}{R_v^3-1},
\qquad \tau_{\rm trunc}=\frac{\mu_0\sigma a^2}{k^2}.
\end{equation}
The measured decay times for $\mathbf A_{\rm ind}=0$ at $R_v$ agree with this prediction to within $0.02\%$ to $0.14\%$. 

\subsection{Sphere dynamo}\label{sec:dynamo}

The sphere dynamo tests coupled field growth against an independent eigenvalue problem for a conducting sphere surrounded by vacuum. In every conductor cell we prescribe the $s_2t_2$ flow of Dudley and James~\cite{DudleyJames89}, named after its poloidal ($s$) and toroidal ($t$) parts of degree two,
\begin{equation}\label{eq:dj-flow}
\mathbf u=R_m\left(\mathbf t_2^0+\epsilon\,\mathbf s_2^0\right),\qquad
\mathbf t_2^0=\nabla\times\left[F(r)\,Y\hat{\mathbf r}\right],\qquad
\mathbf s_2^0=\nabla\times\nabla\times\left[F(r)\,Y\hat{\mathbf r}\right].
\end{equation}
Here $r\leq1$ is the radius in units of $a$, $\theta$ is the polar angle, and $\hat{\mathbf r}$ is the radial unit vector. The angular factor $Y=P_2(\cos\theta)=(3\cos^2\theta-1)/2$ is the unnormalized axisymmetric harmonic of degree $l=2$ and order $m=0$, which the subscript and superscript of $\mathbf t_2^0$ and $\mathbf s_2^0$ denote. Both parts share the radial profile $F(r)=r^2\sin(\pi r)$, the constant $\epsilon=0.14$ weights the poloidal part, and the magnetic Reynolds number $R_m$ sets the amplitude. The flow is impermeable but not no-slip at $r=1$, so the insulating condition on $\phi$ includes the motional term. The DtN map acts directly on the conductor surface, without a vacuum mesh, on the $N16$, $N24$, and $N36$ meshes. The growing mode has $m=1$.

The independent $m=1$ spectral reference, which uses a poloidal-toroidal expansion in harmonics of order $m=1$, parity-folded Chebyshev collocation in radius, and a dense eigensolver~\cite{cpcVerificationData}, gives $R_c=53.562588$ and the growth rate $\lambda(70)=3.6183558$ at $R_m=70$. Its threshold lies $0.000188$ above the value tabulated by Ivers~\cite{IversPRSA17}.

With growth rates fitted over $1.0\leq t\leq1.2$, the computed thresholds are $54.0367$, $53.7748$, and $53.6574$ on $N16$, $N24$, and $N36$, or $0.885\%$, $0.396\%$, and $0.177\%$ above the spectral value, with an observed order of $1.98$. The Richardson consistency estimates are $53.5621$ at this order and $53.5635$ at order two, which differ from the reference by $-0.0005$ and $+0.0009$.

With the start-of-step predictor of Section~\ref{sec:problem} at $n_c=20$, the growth rate $\lambda(70)$ lies $1.423\%$, $0.525\%$, and $0.202\%$ below the spectral growth rate on the three meshes, with an observed order of $2.52$. The Richardson estimates, $3.6176$ and $3.6204$, differ from the reference by $-0.020\%$ and $+0.056\%$, respectively. Omitting the predictor biases off-threshold growth rates by about $1\%$ at $n_c=20$ but leaves the threshold essentially unchanged, since repeating the $N24$ bracket with the predictor shifts it by $+0.0006$.

\begin{figure}[!htbp]
\centering
\includegraphics[width=\linewidth]{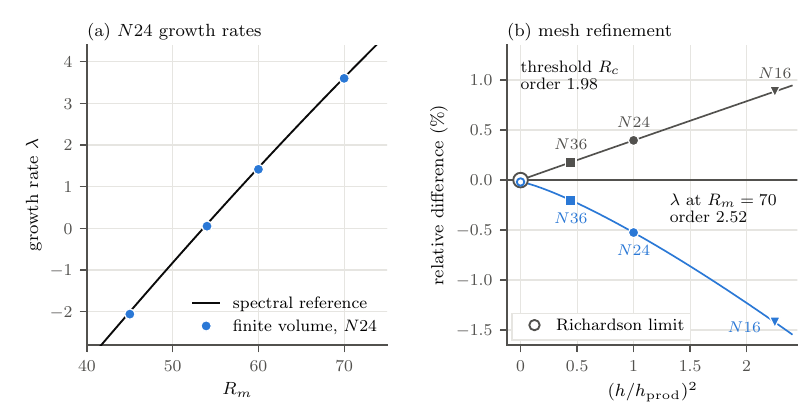}
\caption{Sphere $m=1$ dynamo against the spectral reference. (a) Spectral growth-rate curve and $N24$ growth rates at $R_m=45$, $54$, $60$, and $70$. (b) Relative difference from the reference against $(h/h_{\rm prod})^2$, where $h_{\rm prod}$ is the $N24$ mesh spacing. Growth rates use the predictor, and $R_c$ is computed without it.}
\label{fig:dynamo}
\end{figure}

On the production mesh, the growth rates at $R_m=45$, $54$, $60$, and $70$ lie $0.055$ to $0.019$ below the spectral curve, consistent with the threshold on that mesh lying $0.21$ above the spectral threshold. Figure~\ref{fig:dynamo} shows the growth-rate and refinement comparisons.

On the same mesh, raising $L_{\rm SH}$ from $8$ to $12$ shifts the threshold by $-0.002$, and halving $\Delta t$ shifts it by $+0.0007$. Both shifts are at least forty times smaller than the finest-mesh difference of $0.095$ from Ivers' threshold. Lowering $L_{\rm SH}$ to $6$ shifts the threshold by $-0.020$, and lowering it to $4$ shifts it by $+0.56$. Adding a meshed vacuum buffer out to $R_v=1.5$, so that the map acts on the sphere $r=1.5$ instead of on the conductor surface, raises $\lambda(70)$ by $0.15\%$. The outer condition matters far more than these settings. In shorter runs with this buffer and without the predictor, with growth rates fitted over $0.4\leq t\leq0.6$, $\lambda(70)$ lies $1.4\%$ below the spectral growth rate with DtN on $\Gamma$ and $35\%$ above it with pseudo-vacuum on $\Gamma$, whereas close truncation turns growth into decay, a growth-rate error of $-107\%$ over $0.3\leq t\leq0.5$. Changing the relaxation factor $\gamma$ of Eq.~\eqref{eq:relaxation} from $0.1$ to $0.05$ or $0.2$ moves the $N24$ threshold by at most $1.5\times10^{-4}$, whether the growth rates are fitted over $0.8\leq t\leq1.0$ or $1.0\leq t\leq1.2$. In runs without the predictor at $R_m=70$, the DtN mismatch of Section~\ref{sec:protocol}, sampled at the end of every 500th time step, stays below $1.3\times10^{-4}$ on all three meshes.

Because the coupled trace is not Coulomb compatible, this test also involves the non-solenoidal family. At $R_m=70$ with the predictor, the $l=1$ compatibility residual of Eq.~\eqref{eq:compatibility}, normalized as $\|\alpha_{1m}+2\beta_{1m}\|/(2\|\beta_{1m}\|)$ with each norm taken over the orders $m=-1$, $0$, and $1$, is $0.0446$, $0.0212$, and $0.0095$ on $N16$, $N24$, and $N36$, decreasing at about second order. With a shell at $R_v=1.5$ it is $0.057$ on $N24$. By Eq.~\eqref{eq:third-coeff}, this residual measures the non-solenoidal content of the trace rather than the convergence of Algorithm~\ref{alg:boundary}. The residual thus vanishes under refinement at about the same order as the threshold error, and the growth rates and threshold still converge toward the spectral reference (Fig.~\ref{fig:dynamo}(b)), so this content does not limit the accuracy of the test.

\subsection{Oblate onset}\label{sec:onset}

The oblate dynamo tests the configuration that Ref.~\citenum{FournierECCOMAS06} proposed but did not test, namely a non-spherical conductor in a meshed buffer with the DtN map on an enclosing sphere. Following Ivers~\cite{IversPRSA17}, we write an axisymmetric flow in cylindrical coordinates $(s,\varphi,z)$ as $\mathbf u=\nabla\times(\psi\hat{\boldsymbol\varphi}/s)+s\,\omega\,\hat{\boldsymbol\varphi}$, with stream function $\psi$ and angular velocity $\omega$. The map $\psi(s,z)=ac\,\psi_0(s/a,z/c)$, $\omega(s,z)=\omega_0(s/a,z/c)$ takes the sphere flow $(\psi_0,\omega_0)$ of Eq.~\eqref{eq:dj-flow} to a spheroid with equatorial semi-axis $a$ and polar semi-axis $c$. In units of $a$, the sphere flow has
\begin{equation}\label{eq:oblate-flow}
\psi_0(s,z)=3\epsilon R_m\frac{s^2z\sin(\pi r)}{r},\qquad
\omega_0(s,z)=3R_m\frac{z\sin(\pi r)}{r},\qquad r=\sqrt{s^2+z^2},
\end{equation}
with $\epsilon=0.14$ as in Section~\ref{sec:dynamo}. This flow is incompressible and tangent to the spheroid, and Ivers' definition of $R_m$ is the same as ours. We compress the sphere meshes by $z\mapsto(c/a)z$ and move the shell nodes along radial columns so that $\Gamma$ remains a sphere at $R_v=1.5$.

Growth rates are fitted over $1.0\leq t\leq1.2$, and each threshold follows from three bracketing growth rates as in Section~\ref{sec:protocol}. Fitting over $0.6\leq t\leq0.8$ or $0.8\leq t\leq1.0$ instead moves each threshold in Table~\ref{tab:onset} by less than $0.011$, and by less than $0.0006$ for the later window. The thresholds were computed without the predictor (Table~\ref{tab:settings}). For $a/c=1.5$ on $N24$, repeating the bracket with the predictor shifts the threshold by $-0.0004$. Sampled at the end of every 100th step between $t=1.2$ and $1.3$, the DtN mismatch of Section~\ref{sec:protocol} stays below $1.7\times10^{-5}$.

\begin{figure}[!htbp]
\centering
\includegraphics[width=\linewidth]{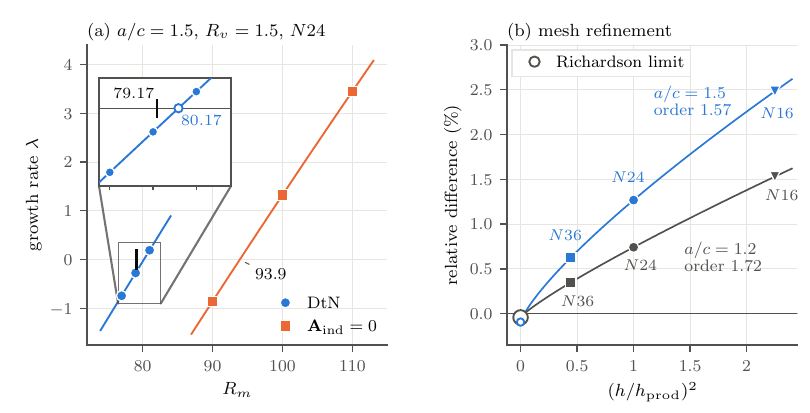}
\caption{Oblate dynamo onset at $R_v=1.5$. (a) Growth rates and quadratic fits at $a/c=1.5$ on $N24$ for DtN ($R_m=77$, $79$, $81$) and $\mathbf A_{\rm ind}=0$ ($R_m=90$, $100$, $110$). The inset enlarges the DtN zero crossing at $80.17$ near Ivers' value $79.1692$ (tick). The $\mathbf A_{\rm ind}=0$ crossing at $93.9$ is marked. (b) Threshold difference from Ivers' value, Eq.~\eqref{eq:reldiff}, against $(h/h_{\rm prod})^2$.}
\label{fig:onset}
\end{figure}

At both shapes, the thresholds approach Ivers' references under three-mesh refinement (Table~\ref{tab:onset}). For $a/c=1.5$, the observed order of $1.57$ gives a Richardson estimate $-0.094\%$ from Ivers' value, and the order-two estimate is $+0.112\%$. For $a/c=1.2$, the observed order of $1.72$ gives $-0.042\%$, and order two gives $+0.034\%$. The two estimates straddle each tabulated reference, which we take as a consistency check rather than as a bound on the discretization error. Figure~\ref{fig:onset} shows both series.

\begin{table}[!htbp]
\centering
\caption{Oblate onset with DtN at $R_v=1.5$, except for the marked radius check. Errors are relative to Ivers' spectral thresholds, and the Richardson rows are consistency estimates.}
\label{tab:onset}
\small
\begin{tabular}{@{}clcc@{}}
\hline
$a/c$ & mesh or estimate & $R_c$ & error \\
\hline
$1.5$ & $N16$ & $81.1352$ & $+2.483\%$ \\
 & $N24$ & $80.1732$ & $+1.268\%$ \\
 & $N24$, $R_v=2$ & $80.1967$ & $+1.298\%$ \\
 & $N36$ & $79.6647$ & $+0.626\%$ \\
 & observed order $1.57$ & $79.0946$ & $-0.094\%$ \\
 & order two & $79.2580$ & $+0.112\%$ \\
 & Ivers & $79.1692$ & \\
\hline
$1.2$ & $N16$ & $63.4486$ & $+1.530\%$ \\
 & $N24$ & $62.9558$ & $+0.741\%$ \\
 & $N36$ & $62.7101$ & $+0.348\%$ \\
 & observed order $1.72$ & $62.4660$ & $-0.042\%$ \\
 & order two & $62.5136$ & $+0.034\%$ \\
 & Ivers & $62.4924$ & \\
\hline
\end{tabular}
\end{table}

The radius check gives $R_c=80.1967$ at $R_v=2$, a shift of $+0.029\%$ from $R_v=1.5$ on the same $a/c=1.5$, $N24$ conductor mesh. Estimated from the growth-rate change at $R_m=79$ and the local slope, rather than from new brackets, halving $\Delta t$ shifts this threshold by $-0.0007$, and raising $L_{\rm SH}$ to $12$ leaves it unchanged to four decimal places. Both changes are below $1\%$ of the change of $0.51$ from $N24$ to $N36$.

With the predictor, the $N24$ bracket gives $80.1728$ at both $n_c=20$ and $5$, and the run with $n_c=5$ takes about a third of the wall time. Its growth rates are, however, $0.8\%$ to $1.1\%$ low, so off-threshold growth rates require $n_c\geq10$ in the tested cases.

Close truncation changes the onset substantially. Imposing $\mathbf A_{\rm ind}=0$ on the same sphere at $R_v=1.5$ gives $R_c=93.895$, $17.1\%$ above the DtN value on the same mesh (Fig.~\ref{fig:onset}(a)). This result shows only that truncation this close to the conductor is inadequate here. Section~\ref{sec:cost} compares DtN with the same condition on a graded shell that extends farther outward.

\subsection{Braking rotor}\label{sec:rotor}

The braking rotor tests two-way coupling between the magnetic field and the rotation against an exact transient reference. A rigid conducting sphere rotates about the $z$ axis in the applied field $\mathbf B_{\rm ext}=B_0\hat{\mathbf x}$. With $\mu_0\sigma a^2=1$, the angular speed $\Omega$ of the rotor equals $R_m$, and the initial speeds are $\Omega_0=R_{m0}=1$, $10$, and $30$. The Ohmic cell currents $\mathbf j_c$ of Eq.~\eqref{eq:ohm} give the torque and the dipole moment, and the speed is then updated explicitly,
\begin{equation}\label{eq:rotor-coupling}
\begin{aligned}
T_z&=\left[\sum_c V_c\,\mathbf r_c\times(\mathbf j_c\times\mathbf B_c)\right]_z,
\qquad \mathbf m=\frac12\sum_c V_c\,\mathbf r_c\times\mathbf j_c,\\
\Omega^{n+1}&=\Omega^n+\frac{\Delta t}{I}T_z^n,\qquad I=\frac{8\pi\rho a^5}{15}.
\end{aligned}
\end{equation}
Here $V_c$, $\mathbf r_c$, and $\mathbf B_c$ are the volume, centroid position, and magnetic field of cell $c$, $n$ is the time-step index, and $I$ is the moment of inertia of the sphere of density $\rho$. We impose $\mathbf u=\Omega\hat{\mathbf z}\times\mathbf r$ in the cells and on the wall faces, so this rigid-body calculation solves no fluid momentum equation. In the rotating frame, the applied field drives pure diffusion, and only $l=1$ is excited. \ref{app:rotor} gives the modal reference and its bounded variants. At $t=0$ the applied field fills the sphere with $\mathbf A_{\rm ind}=0$ and is not ramped. The densities $\rho/(\mu_0\sigma^2B_0^2a^2)=1.5$, $1$, and $0.33$ for $R_{m0}=1$, $10$, and $30$ give reference half-speed times of $4.21$, $4.16$, and $4.16$.

Five exterior treatments share one $N24$ conductor mesh. They are DtN at $R_v=1.5$ and $2$, a radially graded shell with $\mathbf A_{\rm ind}=0$ at its outer radius $R_v=15$, and pseudo-vacuum and $\mathbf A_{\rm ind}=0$, both at $R_v=1.5$. The matched-$\Omega$ torque error is the mean, over $t>1$, of the relative difference from the reference torque at the run's current $\Omega$. Where needed, the reference is continued past $t=5$. Table~\ref{tab:rotor} also gives the relative speed error at $t=5$.

\begin{table}[!htbp]
\centering
\caption{Braking rotor errors against the transient reference. Torque errors are matched-$\Omega$ means over $t>1$, and speed errors are taken at $t=5$.}
\label{tab:rotor}
\small
\setlength{\tabcolsep}{6pt}
\begin{tabular}{@{}clrr@{}}
\hline
$R_{m0}$ & exterior & $T_z$ error & $\Omega(5)$ error \\
\hline
$1$ & DtN $R_v=1.5$ & $-0.393\%$ & $+0.328\%$ \\
 & DtN $R_v=2$ & $-0.393\%$ & $+0.328\%$ \\
 & graded $R_v=15$ & $-0.392\%$ & $+0.328\%$ \\
 & pseudo-vacuum $R_v=1.5$ & $-0.315\%$ & $+0.442\%$ \\
 & $\mathbf A_{\rm ind}=0$, $R_v=1.5$ & $-0.578\%$ & $+0.125\%$ \\
\hline
$10$ & DtN $R_v=1.5$ & $-0.332\%$ & $+0.358\%$ \\
 & DtN $R_v=2$ & $-0.334\%$ & $+0.360\%$ \\
 & graded $R_v=15$ & $-0.363\%$ & $+0.395\%$ \\
 & pseudo-vacuum $R_v=1.5$ & $-6.012\%$ & $+7.093\%$ \\
 & $\mathbf A_{\rm ind}=0$, $R_v=1.5$ & $+9.518\%$ & $-11.297\%$ \\
\hline
$30$ & DtN $R_v=1.5$ & $-0.385\%$ & $+0.691\%$ \\
 & DtN $R_v=2$ & $-0.390\%$ & $+0.700\%$ \\
 & graded $R_v=15$ & $-0.469\%$ & $+0.841\%$ \\
 & pseudo-vacuum $R_v=1.5$ & $-14.481\%$ & $+25.146\%$ \\
 & $\mathbf A_{\rm ind}=0$, $R_v=1.5$ & $+28.657\%$ & $-51.391\%$ \\
\hline
\end{tabular}
\end{table}

The DtN torque changes little between the two tested radii. Their matched-$\Omega$ errors differ by at most $0.006$ percentage points in every $R_m$ bin for all three initial speeds.

The graded shell stays close to DtN in torque, with mean differences of $0.03$ points at $R_{m0}=10$ and $0.08$ at $30$. The predicted error of $\mathbf A_{\rm ind}=0$ at $R_v=15$, from the bounded model of \ref{app:rotor}, is $+0.007\%$ to $+0.035\%$, opposite in sign to the measured difference and about three times smaller. In these tests the difference therefore comes from the shell discretization.

\begin{figure}[!htbp]
\centering
\includegraphics[width=\linewidth]{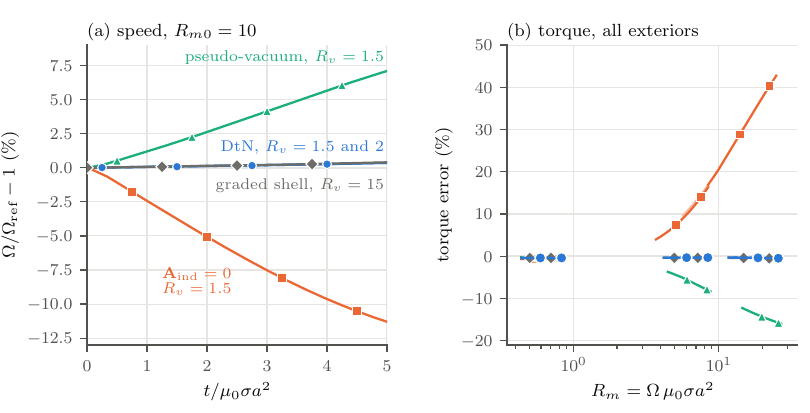}
\caption{Braking rotor against the transient reference on $N24$. (a) Speed error $\Omega(t)/\Omega_{\rm ref}(t)-1$ at $R_{m0}=10$. (b) Matched-$\Omega$ torque error against $R_m$ over $1\leq t\leq5$ for $R_{m0}=1$, $10$, and $30$, with the line styles of (a). DtN at $R_v=2$ (dashed) coincides with DtN at $R_v=1.5$ (solid). The light orange line is the late part of the $R_{m0}=30$ run with $\mathbf A_{\rm ind}=0$, which overlaps the $R_{m0}=10$ run.}
\label{fig:rotor}
\end{figure}

The two local conditions at $R_v=1.5$, by contrast, carry substantial exterior modeling errors at finite $R_m$. At $R_{m0}=10$, pseudo-vacuum and $\mathbf A_{\rm ind}=0$ miss the reference torque by $-6.01\%$ and $+9.52\%$, and at $30$ by $-14.48\%$ and $+28.66\%$ (Fig.~\ref{fig:rotor}). Their differences from DtN in the same $R_m$ bin match the modeling errors predicted by the bounded models of \ref{app:rotor} to within $0.46$ points. At $R_{m0}=1$, all exterior treatments are within $0.25$ points of DtN, because the induced field is $O(R_m)$ relative to the applied field.

Under mesh refinement, the DtN torque error relative to the exact transient reference decreases. At $R_{m0}=10$ and $R_v=1.5$, it is $-0.7521\%$, $-0.3315\%$, and $-0.1425\%$ on $N16$, $N24$, and $N36$. The observed order of $1.97$ gives a Richardson consistency estimate of $+0.012\%$, and order two gives $+0.009\%$.

The speed trajectory follows from the computed torque. At $t=5$, the $\Omega$ difference equals the time-integrated torque difference divided by $I$ to within $1.5\%$ of that difference, which links the speed errors in Table~\ref{tab:rotor} to the torque differences.

\section{Accuracy and cost}\label{sec:cost}

This section weighs accuracy against cost in two of the intended uses, a non-spherical conductor inside a meshed buffer (oblate onset, Section~\ref{sec:onset}) and a field coupled to the motion of a rigid conductor (the braking rotor, Section~\ref{sec:rotor}). On the outer sphere at $R_v=1.5$, the local conditions carry large errors. Close truncation raises the oblate threshold by $17.1\%$, and pseudo-vacuum and close truncation miss the rotor torque at $R_{m0}=30$ by $-14.48\%$ and $+28.66\%$. The condition $\mathbf A_{\rm ind}=0$ carries an $O(1)$ error at its own boundary, but the dipole part of this error is reduced in proportion to $(a/R_v)^3$ by the time it reaches the conductor. Moving the boundary outward therefore makes $\mathbf A_{\rm ind}=0$ competitive, and radial grading keeps the larger radius cheap. We compare DtN at $R_v=1.5$ with $\mathbf A_{\rm ind}=0$ on a graded shell to $R_v=15$. The shell has the same angular mesh and first radial cell size, $0.025$, as the DtN buffer, and its layers grow by a factor of $1.077$, giving $51$ layers. The oblate shell is mapped along radial columns as in Section~\ref{sec:onset}, which retains the grading. In the rotor, the predicted error of $\mathbf A_{\rm ind}=0$ at $R_v=15$ is $+0.007\%$ to $+0.035\%$ (Section~\ref{sec:rotor}), so the comparison tests cost at nearly matched accuracy.

Under refinement, the DtN boundary discretization error vanishes at fixed $\Gamma$, whereas the $\mathbf A_{\rm ind}=0$ error of the graded shell is set by its $R_v$. If the error is dominated by the dipole, the radial grading is geometric, and the discretization is second order, keeping the $\mathbf A_{\rm ind}=0$ error level with the grid error requires the graded shell's $R_v\propto h^{-2/3}$, where $h$ is the mesh spacing. Because geometric grading makes the layer count grow as $\ln R_v$, the predicted cell advantage of DtN grows as $\ln(1/h)$. This is a scaling argument under those assumptions rather than a measurement, and the observed oblate orders are $1.57$ and $1.72$. The $l=1$ vector potential of the non-solenoidal family decays only as $r^{-1}$, yet the graded and DtN oblate thresholds below differ by only $0.006\%$.

\begin{table}[!htbp]
\centering
\caption{Matched comparisons on $N24$. The result column gives $R_c$ or the matched-$\Omega$ torque error. Oblate onset uses the predictor, which gives $80.1728$ rather than the $80.1732$ of Table~\ref{tab:onset}, and $32$ ranks per case in bundles of three, with the bundle-mean wall time. Rotor rows are for $R_{m0}=30$, with $16$ ranks per case and the five exteriors of Table~\ref{tab:rotor} run concurrently on one node.}
\label{tab:cost}
\small
\setlength{\tabcolsep}{4pt}
\begin{tabular}{@{}llrcrc@{}}
\hline
test & exterior & cells & $n_c$ & result & wall (s) \\
\hline
oblate onset & DtN $R_v=1.5$ & $69\,120+114\,048$ & $20$ & $80.1728$ & $3765$ \\
 & graded $R_v=15$ & $69\,120+176\,256$ & $20$ & $80.1774$ & $3835$ \\
 & DtN $R_v=1.5$ & $69\,120+114\,048$ & $5$ & $80.1728$ & $1257$ \\
 & graded $R_v=15$ & $69\,120+176\,256$ & $5$ & $80.1778$ & $1450$ \\
\hline
rotor & DtN $R_v=1.5$ & $69\,120+69\,120$ & $20$ & $-0.385\%$ & $7043$ \\
 & graded $R_v=15$ & $69\,120+176\,256$ & $20$ & $-0.469\%$ & $8117$ \\
 & pseudo-vacuum $R_v=1.5$ & $69\,120+69\,120$ & $20$ & $-14.481\%$ & $6906$ \\
 & $\mathbf A_{\rm ind}=0$, $R_v=1.5$ & $69\,120+69\,120$ & $20$ & $+28.657\%$ & $6631$ \\
\hline
\end{tabular}
\end{table}

In oblate onset the two thresholds differ by $0.006\%$ (Table~\ref{tab:cost}). DtN uses $25\%$ fewer cells and $13\%$ less wall time at $n_c=5$, but only about $2\%$ less wall time at $n_c=20$. In the rotor at $R_{m0}=30$, the two torque errors differ by $0.08$ percentage points, and DtN uses $44\%$ fewer cells and $13\%$ less wall time. At $R_{m0}=1$ and $10$, the graded-shell wall time is $1.22$ and $1.17$ times that of DtN. On the same mesh, pseudo-vacuum and close truncation save only $2\%$ to $3\%$ and $6\%$ to $11\%$ of the DtN wall time over the three initial speeds. In the sphere dynamo, the map acts on the conductor surface itself, so the exterior adds no cells (Section~\ref{sec:dynamo}). Operator evaluation is a small part of the cost. In the $a/c=1.5$ oblate run at $R_m=80$ and $n_c=20$, it takes $2.9\%$ to $4.7\%$ of each time step on $16$ to $128$ ranks~\cite{cpcVerificationData}.

\section{Limitations}\label{sec:limits}

The computed vector potential is not exactly divergence-free. This discrete gauge error arises in the regional solver, which does not impose the Coulomb gauge (Section~\ref{sec:problem}), rather than in the DtN closure, which only continues the trace it receives, including its non-solenoidal part (Section~\ref{sec:exactness}). In the sphere dynamo at $R_m=70$, the bulk ratio $\|\nabla\cdot\mathbf A\|/\|\nabla\times\mathbf A\|$ of volume-weighted $L^2$ norms over the conductor is about $2\%$, namely $2.2\times10^{-2}$ on $N24$ and $2.0\times10^{-2}$ on $N36$. It does not change with the predictor or with an added shell, which moves the closure from the conductor surface to $r=1.5$, and it peaks in the conductor core, away from $\Gamma$. This points to a property of the conductor discretization, whose effect on onset we have not quantified. Imposing the Coulomb gauge in the regional solver could address this error at its source, and we leave this to future work.

This solver-side gauge error also separates two definitions of the current. For a converged solution of the continuum induction equation, the Maxwell current $\nabla\times\mathbf B/\mu_0$ and the Ohmic current of Eq.~\eqref{eq:ohm} differ by $\nabla(\nabla\cdot\mathbf A)/\mu_0$. The discrete ratio below estimates only this gauge term. The full discrete difference also contains the equation residual and operator inconsistency. In the near-onset $a/c=1.5$ runs of Section~\ref{sec:onset} without the predictor, the conductor's $\operatorname{rms}|\nabla(\nabla\cdot\mathbf A)|$ is $3.8\%$, $3.1\%$, and $2.6\%$ of $\operatorname{rms}|\nabla\times\mathbf B|$ on $N16$, $N24$, and $N36$, computed with the least-squares gradients that the solver uses in these runs. Recomputed with Gauss gradients, the same ratio rises from $9.4\%$ to $13.9\%$. Because this second-derivative measure depends on the operator, we claim no convergence for it. The same gauge term leaves a small current in the meshed buffer, where $\nabla^2\mathbf A=0$ makes $\nabla\times\mathbf B$ equal to $\nabla(\nabla\cdot\mathbf A)$ in the continuum. In the near-onset oblate runs, the buffer's $\operatorname{rms}|\nabla\times\mathbf B|$ is $1.5\%$, $1.2\%$, and $0.9\%$ of the conductor's on $N16$, $N24$, and $N36$ with least-squares gradients, and about $1.5\%$ on all three with Gauss gradients. It is largest next to the flattened conductor and at $\Gamma$. We have not quantified its effect on onset separately, but the thresholds still converge toward Ivers' values (Section~\ref{sec:onset}).

Every current-carrying conductor must lie inside the enclosing sphere. A non-spherical conductor needs a meshed buffer that extends to the sphere. In addition, the boundary update is explicit and under-relaxed, which can add solves per time step. Finally, the radius checks compare distinct discretized problems, so they do not establish independence from every enclosing radius.

Free surfaces may lie inside $\Gamma$, but no free-surface response is tested here. Free-surface calculations are left for future work, separate from the prescribed-flow and rigid-body tests reported above.

\section{Conclusions}\label{sec:conclusions}

We have presented a DtN closure for the exterior field at finite $R_m$ on a fixed enclosing sphere and released it as part of the open-source solver \texttt{FreeMHD}~\cite{FreeMHDcode}. It matches each Cartesian component of the vector potential to its decaying exterior solution and thereby retains all three decaying vector-Laplace families. Unlike the spherical map of Ref.~\citenum{FournierECCOMAS06}, which carries only the toroidal family, the closure therefore also applies to a discrete trace that has a radial component and is not Coulomb compatible. Non-spherical conductors are connected to $\Gamma$ through a meshed buffer. On the finest meshes, the oblate onset thresholds differ from Ivers' spectral values by $0.63\%$ and $0.35\%$ at the two shapes, and the Richardson estimates check consistency rather than bound the error.

For the sphere dynamo, growth rates approach an independent $m=1$ spectral solution. The braking rotor tests the feedback of the magnetic torque on the rigid-body speed against an exact transient reference. In the three braking runs with DtN at $R_v=1.5$ on $N24$, the mean matched-$\Omega$ torque error stays below $0.4\%$. At $R_{m0}=10$ and $30$, pseudo-vacuum and close truncation produce larger errors, consistent with their bounded exterior models in \ref{app:rotor}.

The DtN closure keeps its enclosing sphere fixed under refinement, here at $R_v=1.5$. At nearly matched accuracy, it uses $25\%$ and $44\%$ fewer cells than a graded shell extending to $15$ conductor radii in the oblate and rotor tests, and $2\%$ to $18\%$ less wall time in these runs. The coupled results should be interpreted with the discrete gauge and buffer-current limitations of Section~\ref{sec:limits} in mind.

\section*{Acknowledgments}

This research was supported by the U.S. Department of Energy under Award DE-SC0024626, and by R\&D Program of ``Optimal Basic Design of DEMO Fusion Reactor, CN2602-2'' through the Korea Institute of Fusion Energy (KFE) funded by the Government funds. The authors gratefully acknowledge the Research Institute of Energy and Resources and the Institute of Engineering Research at Seoul National University.

\section*{Declaration of competing interest}
The authors declare that they have no known competing financial interests or personal relationships that could have appeared to influence the work reported in this paper.

\section*{CRediT authorship contribution statement}
\textbf{Min Ki Jung:} Conceptualization, Methodology, Software, Validation, Formal analysis, Investigation, Data curation, Visualization, Writing -- original draft, Writing -- review \& editing. \textbf{Brian Wynne:} Investigation, Validation, Writing -- review \& editing. \textbf{Francisco Saenz:} Investigation, Validation, Writing -- review \& editing. \textbf{Yufan Xu:} Investigation, Validation, Writing -- review \& editing. \textbf{Egemen Kolemen:} Conceptualization, Funding acquisition, Supervision, Writing -- review \& editing. \textbf{Yong-Su Na:} Funding acquisition, Supervision, Writing -- review \& editing.

\section*{Declaration of generative AI and AI-assisted technologies in the manuscript preparation process}
During the preparation of this work, the authors used Claude (Anthropic) and Codex (OpenAI) in order to edit the language and structure of the manuscript and to check its internal consistency. After using these tools, the authors reviewed and edited the content as needed and take full responsibility for the content of the published article.

\section*{Data availability}
The boundary module is in the solver \texttt{apotBCMultiRegionInterFoamDtN} of the open-source code \texttt{FreeMHD}~\cite{FreeMHDcode}, distributed under the MIT License. The version used here is tag \texttt{dtn-v1}, commit \texttt{a36d930} on the main branch. The verification inputs, independent references, and analysis data are identified in the archive~\cite{cpcVerificationData}.

\appendix
\section{Exact references for the braking rotor}\label{app:rotor}

Write the applied field in the rotating frame as $A_c(t)=B_0$$e^{-i\chi(t)}$, where the rotor angle $\chi$ satisfies $\dot\chi=\Omega$. Rigid rotation turns induction into diffusion with $\eta=(\mu_0\sigma)^{-1}$. For one transverse direction, we use spherical coordinates $(r,\vartheta,\varphi)$ whose polar axis is that transverse direction and write the vector potential as $g(r,t)\sin\vartheta\,\hat{\boldsymbol\varphi}$, so the radial equation and the insulating matching condition are
\begin{equation}\label{eq:rotor-radial}
\partial_tg=\eta\left(g''+\frac{2g'}r-\frac{2g}{r^2}\right),\qquad
g(0)=0,\qquad g'(a)+\frac{2g(a)}a=\frac32A_c(t).
\end{equation}
The last condition matches $g$ to the exterior solution $A_cr/2+d/r^2$ with dipole coefficient $d$. Complex transverse solutions can be superposed. Setting $g=A_cr/2+w$ makes the condition on $w$ homogeneous. With the spherical Bessel functions $j_0$ and $j_1$, expanding $w=\sum_n c_nj_1(k_nr/a)$ gives $j_0(k_n)=0$, so $k_n=n\pi$ and
\begin{equation}\label{eq:rotor-modes}
\frac ra=\sum_{n=1}^{\infty}b_nj_1(k_nr/a),\qquad
b_n=\frac{6(-1)^{n+1}}{n\pi},\qquad
\dot c_n=-\eta\left(\frac{k_n}{a}\right)^2c_n-\frac a2b_n\dot A_c.
\end{equation}
Here $j_1(k_n)=(-1)^{n+1}/(n\pi)$. The exterior dipole coefficient and the mechanical equation close the reference,
\begin{equation}\label{eq:rotor-ode}
\begin{aligned}
D&=d_x+id_y=a^2\sum_nc_nj_1(k_n),\qquad
T_z=-\frac{4\pi}{\mu_0}\operatorname{Im}(A_c^{*}D),\\
I\dot\Omega&=T_z,\qquad I=\frac{8\pi\rho a^5}{15}.
\end{aligned}
\end{equation}
At $t=0$, $c_n=0$, $\chi=0$, and $\Omega=\Omega_0$.

The reference integrates Eqs.~\eqref{eq:rotor-modes} and~\eqref{eq:rotor-ode} with the density and $\Omega_0$ of each run. The finite-volume result, in contrast, takes the torque from the cell-integrated Lorentz force, so the comparison tests both induction and mechanical feedback.

At fixed speed, bounded exterior solutions isolate the effect of boundary modeling. For a transverse $l=1$ mode with $a=1$, we write the induced exterior scalar potential as $B_0(D_br^{-2}+Cr)\cos\vartheta$, where the unbounded solution has $C=0$ and the growing $Cr$ harmonic is admissible because the exterior is bounded. The boundary condition is imposed at $R_v$, which is $1.5$ for pseudo-vacuum and $\mathbf A_{\rm ind}=0$ and $15$ for the graded shell. Pseudo-vacuum imposes $\mathbf n\times\mathbf B_{\rm ind}=0$ at $R_v$, which gives $C=-D_b/R_v^3$, whereas $\mathbf A_{\rm ind}=0$ gives $B_r=0$ and $C=2D_b/R_v^3$. We define
\begin{equation}\label{eq:rotor-bounded}
\begin{aligned}
\xi&=\sqrt{iR_m},&\mathcal R&=\frac{j_1(\xi)}{\xi j_1'(\xi)},&
q&=\frac{\mathcal R-1}{2(2\mathcal R+1)},\\
D_{\rm pv}&=\frac{q}{1-qR_v^{-3}},&
D_{\mathbf A=0}&=\frac{q}{1+2qR_v^{-3}}.
\end{aligned}
\end{equation}
Matching the potential and the magnetic field at the conductor surface yields the two dipole coefficients. As $R_v$ grows, both tend to the unbounded coefficient $q$, which gives the steady torque $T_z=-(4\pi B_0^2/\mu_0)\operatorname{Im}q$ at fixed $\Omega$. The transient reference approaches this torque to within $5\times10^{-9}$ to $3\times10^{-8}$ for the tested speeds, and at low $R_m$ it matches the limit $-(2\pi/15)\sigma B_0^2\Omega a^5$ to within $2\times10^{-8}$. The reference with 400 modes differs from that with 200 modes by at most $4\times10^{-8}$.

The relative modeling error in the steady torque for either bounded condition is
\begin{equation}\label{eq:rotor-model-error}
\delta_{T,\rm bc}=\frac{\operatorname{Im}(D_{\rm bc})}{\operatorname{Im}(q)}-1.
\end{equation}
This quasi-steady estimate is evaluated at the current $R_m$ and compared, bin by bin, with the computed torque difference from DtN. The modal transient remains the reference for braking because its amplitudes retain magnetic memory.

\bibliographystyle{elsarticle-num}
\bibliography{reference}

\end{document}